# Advancements in high refractive index media: from quantum coherence in atomic system to deep sub-wavelength coupling in metamaterials

**Leena Singh, and Weili Zhang***

*School of Electrical and Computer Engineering, Oklahoma State University, Stillwater, Oklahoma 74078, USA*

*\*Corresponding author: weili.zhang@okstate.edu*

Refractive index enhancement is crucial in the field of lithography, imaging, optical communications, solar devices and many more. We present a review of advancements in the process of designing high refractive index metamaterials, starting from quantum coupling and photonic bandgap materials to metamaterials utilizing deep subwavelength coupling to achieve ever-high values of refractive index. A crisp and critical impression of each scheme are presented. The understanding of evolution of material design from intrinsic electronic states manipulation to meta-atoms design is not only fascinating but prerequisite to developing successful devices and applications.

doi:10.3788/COLXXXXXX.XXXXXX.

Materials with modified refractive index have been long part of human history. An impressive application of modified refractive index can be found in the form of well-known stained glass art window at 13th century Notre Dame Cathedral in Paris. Refractive index is basically defined as a dimensionless quantity, which measures the bending of the electromagnetic wave while passing from one medium to another, as shown in Fig. 1.

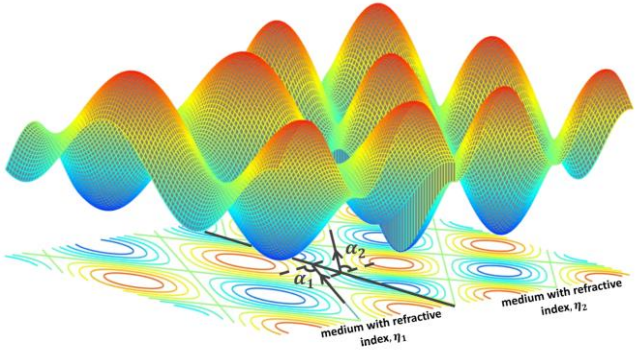

Fig. 1. Bending to an electromagnetic wave while passing from one medium to another with different refractive indices. An electromagnetic wave travelling at an angle of $\alpha_1$ w.r.t. the normal in medium with refractive index $\eta_1$, undergoes refraction at the medium interface and travels at an angle of $\alpha_2$ w.r.t. the normal in medium with refractive index $\eta_2$.

In 1621, Dutch astronomer and mathematician Willebrord Snell defined the relation between the ratio of refractive indices of the two media and ratio between the angle of incidence and angle of refraction as:

$$\frac{\eta_1}{\eta_2} = \frac{\sin\alpha_2}{\sin\alpha_1}. \quad (1)$$

By 19th century, a number of scientific journals had already been discussing the potential applications of high refractive index media for laser particle acceleration[1,2], optical microscopy[3], atomic tests of electroweak physics[4] and magnetometry[5]. In 1991, O. Scully presented the possibility of achieving high refractive index without high absorption via quantum coherence, in principle for the first time[5]. Index enhancement was accomplished by choosing to operate near an atomic resonance between a coherent ground-state doublet configuration |b⟩ and |b'⟩ and an excited state |a⟩ with level probabilities $\rho_{\alpha\alpha}$, $\alpha$ = a, b, c along with an appropriately chosen atom-field detuning[5]. The result is shown in Fig. 2, with large dispersion marked as point B and a near zero absorption marked as point C.

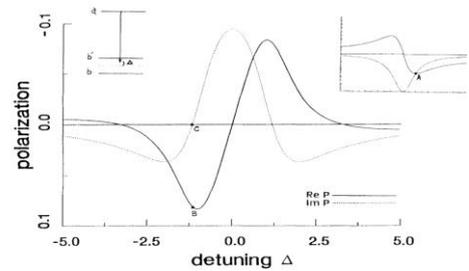

Fig. 2. Dispersive (ReP) and absorptive (ImP) parts of polarization vs detuning of radiation frequency from midpoint between level b' and b. Polarization plotted on arbitrary scale, and detuning $\Delta$, plotted in units of atomic decay. Inset, upper right-hand corner: Usual dispersion-absorption curve. Inset, upper left-hand corner: Present level scheme. [Reprinted with permission from "Scully, M.O., Enhancement of the index of refraction via quantum coherence. Phys. Rev. Lett. 1991, 67(14): p. 1855-1858". Copyright (1991) by the American Physical Society].

Coherence between levels b' and b can be achieved by a number of techniques, i.e., pulse[6], microwave[7] or Raman technique[8,9]. In 1992, M. Fleischhauer, et al. presented a detailed survey of various schemes in which atomic coherence and interference effects lead to complete absorption cancellation and an ultrahigh index of refraction[10].

In 1994, J. Dowling, et al. proposed photonic bandgap materials as a possible candidate to achieve high index of refraction and presented a quantitative study explaining the anomalous index of refraction in photonic band gap materials[11]. The term photonic band gap was introduced for the first time, in 1979 by K. Ohataka[12] when he established the theory of Low Energy Photon Diffraction(LEPD) in analogy with the theory of Low Energy Electron Diffraction(LEED). However, it was in 1990, when the first experimental three-dimensional (3D) photonic band gap made in a periodic dielectric structure was demonstrated by K. M. Ho, et al.[13]. J. Dowling's study was based on the one-dimensional (1D) periodic dielectric array that was used to model a 3D photonic band gap material, to enhance the effective refractive index. He explained that near the gap in the photonic band gap materials the effective index of refraction can become less than unity and in fact can approach zero at the band edge itself-leading to ultra-refractive optical effect[11]. Given that refractive index of a system is the ratio $n_1/n_2$, his scheme was based on decreasing $n_2$ instead of increasing $n_1$, to obtain high effective refractive index[11]. In 1996, Zibrov, et al presented proof-of-principle experiment demonstrating resonantly enhanced refractive index and reduced absorption achieved by quantum coherence in Rb vapor[14]. A resonant change of about $10^{-4}$ was observed in the refractive index.

By the end of 1999, advancements in microfabrication techniques have added ease to the fabrication of artificial media. In addition to 3D structures, two-dimensional (2D) thin film and structures with metallic components within the unit cells, metallo-dielectric structures have also begun to attract scientific attention, as the physics associated with the metallo-dielectric structures was quite unique and very different from the pure dielectric structures[15]. In 2001, the term metamaterial was first defined by R.M. Welser as "macroscopic composites having a man-made, 3D, periodic cellular architecture designed to produce an optimized combination, not available in nature, of two or more responses to a specific excitation. Each cell contains metaparticles, macroscopic constituents designed with low dimensionality that allow each component of the excitation to be isolated and separately maximized. The metamaterial architecture is selected to strategically recombine local quasi-static responses, or to combine or isolate specific non-local responses"[16].

In 2005, J.T. Shen, et al. explained the mechanism for designing a high refractive index metallic metamaterial. Shen claimed that a metallic film with periodic 1D cut through slits, can be considered as a dielectric slab with an effective refractive index, n and width, L, as shown in Fig. 3[17].

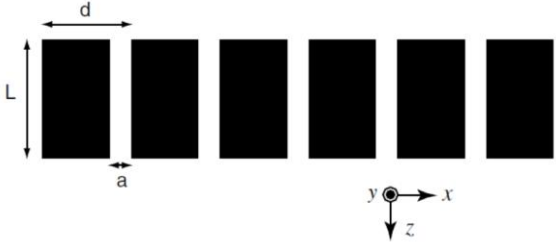

Fig. 3. Schematic of the metal film with periodic slits. The parameters are defined as in the figure: *a* is the width of the slit, d is the periodicity, and L is the thickness of the metal film. The black regions indicate the metal parts, and the white regions are the vacuum. The film is extended in the *x-y* plane. [Reprinted with permission from "Shen, J.T., P.B. Catrysse, and S. Fan, Mechanism for designing metallic metamaterials with a high index of refraction. Phys. Rev. Lett. 2005, 94(19): p. 197401". Copyright (2005) by the American Physical Society].

The effective "n" is determined by field comparison between the metallic film with periodic cut through the slits and the dielectric slab. Given the periodicity of the slits in the metallic film is smaller than the wavelength, the zeroth order transmission amplitude of a TM wave normally incident on the metallic film can be written as[17],

$$t_0 = \frac{4[(f/\emptyset^2)/(1+1/\emptyset^2)]e^{-i(\omega/c)L}}{1-[(1-1/\emptyset)/(1+1/\emptyset)]^2 e^{2i(\omega/c)L}}, \quad (2)$$

where, $f = a/d$ is the filling factor of the slit, and $\emptyset = \sum_{p=-\infty}^{\infty} f g_0^2 \frac{w/c}{\alpha_0}$, for normal incidence, $g_0 = 1$ and $\alpha_0 = w/c$. When a plane is normally incident on a dielectric slab with refractive index *n*, the transmission amplitude can be written as:

$$t = \frac{4[n/(1+n^2)]e^{-in(\omega/c)L'}}{1-[(1-n)/(1+n)]^2 e^{2in(\omega/c)L'}}. \quad (3)$$

Comparing equations (2) and (3), both transmission amplitudes $t_0$ and $t$, can be approximately equalized by setting $n = 1/\emptyset$ and $L' = L/n$ [17]. With the transmission equivalence set up and establishing effective refractive index as $n = d/a$, which is defined only by the geometry of the structure. Shen, et al. pointed that refractive index of a media can be enhanced without having to change the intrinsic electronic states of the material. The paradigm of high refractive index media design is now shifting towards controlling the behavior of the electromagnetic waves by virtue of structural geometry of the artificial media, so as to achieve desired materials parameters namely, permeability and permittivity rather than meddling with the atomic states of the media.

This range of frequencies from 1 to 3 THz is extremely fascinating as it witnesses the breakpoint of both electrical and magnetic response in most of the naturally occurring materials. Terahertz regime is the slice of spectrum that lies between infrared and microwave radiations. In spite of the fact that, terahertz technology had very diverse and attractive applications in the field of astronomy, semiconductor, medical-imaging, atmospheric studies, space-communication, defense and so on, it remains the least developed region due to lack of materials responding to at these frequencies[18,19]. The advent of metamaterials has proven to be especially valuable in the terahertz regime and expected to close the "THz gap". Following the same structure of metallic film with periodically arranged cut through slits, theoretically conversed by J.T. Shen, et al., A. Pimenov and A. Loidl performed first experimental demonstration for a high refractive index metamaterial at terahertz frequencies in 2006. A terahertz high refractive index metamaterials was fabricated with high purity copper plates with periodic slits[20], Mach-Zehnder interferometer arrangement[21] was then used to carry out the experiment for the frequency range 60 to 380 GHz, and reflectance $R = |r|^2$ and complex transmittance $\sqrt{T}e^{i\emptyset}$ were measured. The effective parameters of the sample were obtained using Fresnel optical formulas for reflectance and transmittance, using following equations:

$$t = \frac{(1-r_0^2)t_1}{1-r_0^2 t_1^2}, \quad (4)$$

$$r = \frac{(1-t_1^2)r_0}{1-r_0^2 t_1^2}. \quad (5)$$

Here, $r_0 = (n-1)/(n+1)$ and $t_1 = \exp(2\pi i n L/\lambda)$, $r_0$ is the reflection amplitude at the air-sample interface, $t_1$ is the "pure" transmission amplitude, *n* is the complex refractive index, *L* is the sample thickness, and $\lambda$ is the radiation wavelength[20]. An effective refractive index 5.51 was obtained and an additional correction $\emptyset_{corr} = 2\pi(L - L_{eff})/\lambda$ was suggested to be added to the experimentally obtained phase, in order to establish the equivalence to the real dielectric slab. This metal-dielectric interface can support surface plasmon waves in the optical regime.

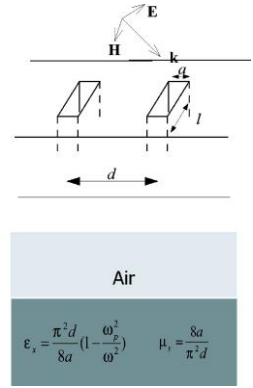

Fig. 4. Equivalent waveguide structure for periodic holes. Metal surface with periodic holes drilled (above) and equivalent waveguide structure (below). [Reprinted with permission from "Z. Ruan and M. Qiu, Slow electromagnetic wave guided in subwavelength region along one-dimensional periodically structured metal surface. Appl. Phys. Lett. 2007, 90(20): p. 201906", with the permission of AIP Publishing]

The dispersion curve of these surface plasmon waves is relatively flat at high frequencies and can be used to slow down light. In the low frequency (microwave) regime, however, metal is treated as perfect electric conductor (PEC), which means electromagnetic waves cannot penetrate into metal and hence metal-dielectric interface cannot support surface waves. So, instead of cut through slits or metal-dielectric interface, an array of holes drilled in metal surface were used to achieve slow electromagnetic waves[22,23]. The schematic and numerical comparison between the metal dielectric structure and metallic hole structure is elaborated in Fig. 4[22].

In 2009, two intriguing structural schemes were simulated to achieve high index of refraction. One was broad bandwidth 3D structure by J. Shin, et al.[24] while another was a resonance based structure by H. Shi, et al.[25]. J. Shin's structure consisted of metal cubes that were arranged in a cubic array fashion, all the six surfaces of these cubes consisted of air slits and were inter-connected by three orthogonal wires, as shown in Fig. 5(a). The structure was based on the observation that the induced charges due to normal electric filed determines effective relative permittivity, $\varepsilon_r$ while the induced surface current due parallel magnetic field contributes to the effective relative permeability, $\mu_r$[24]. For large cube limit (i.e. $b \approx a$ ), permittivity and permeability can be estimated using following equations:

$$\varepsilon_r \approx a/(a-b), \quad (6)$$

$$\mu_r = 1 - b^2/a^2 \approx \frac{2(a-b)}{a}, \quad (7)$$

where 'b' is the side of the cube and 'a' represents the periodicity[24]. The simulated structure was able to enhance the refractive index over a broad bandwidth, by maintaining permeability near unity via reducing the area subtended by the current loops owing to the air slits in the surface and thereby suppressing the diamagnetic response. Meanwhile, the permittivity of the structure is enhanced by upholding a strong capacitive response within the cubic array. The other salient scheme, which was proposed by H. Shi, was based on electric resonance in the surface structures to enhance the effective refractive index[25].

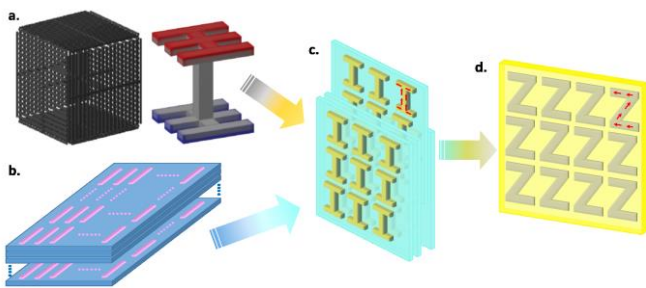

Fig. 5. Metamaterial design evolution. (a) Unit cell of metal cubes that were arranged in a cubic array fashion, all the six surfaces of these cubes consisted of air slits and were inter-connected by three orthogonal wires along with the simplifies structure (left) with two plates with air slits and a connecting wire along z-direction. [Reprinted with permission from "J. Shin, J.T. Shen, and S. Fan, Three-dimensional metamaterials with an ultrahigh effective refractive index over a broad bandwidth. Phys. Rev. Lett. 2009, 102(9): 093903". Copyright (2009) by the American Physical Society]. (b) Layered view of bulk materials formed with unit cell of single cut wire on dielectric substrate. (c) T- shaped metallic patch structure, (d) terahertz metamaterial with Z-shaped meta-atoms.

H. Shi's configuration, as shown in Fig. 5(b) is a planar structure of periodic arrangement of metallic wires on a dielectric substrate of subwavelength thickness, which was then layered to achieve a bulk metamaterial. The value of maximum effective refractive index of the metamaterial is characterized using Bloch wave method and was defined using the following equation:

$$n_{max} = \beta_{max}/k_0 = \pi c/(\omega_0 d), \quad (8)$$

where $\omega_0$ is the resonance angular frequency, $\beta_{max}$ is the propagation constant at the resonance, $k_0$ is the wave number at resonance frequency and $d$ is the periodicity of the structure[25]. The length of the metallic subwavelength wires controlled the resonance frequency while the propagation constant corresponding to higher refractive index around the resonance is governed by the number of layers along the direction of propagation of the electromagnetic wave. This resonance based scheme offered a contrast to the previous prevalent methods which achieved high refractive index by the confinement of electromagnetic mode in the highly subwavelength region, i.e. the propagation TEM mode in the slits and evanescent mode in the holes below the cutoff[25]. This metallic wire array metamaterial design was soon after, experimentally verified and analyzed by X. Wei, et al. in 2010, to achieve high refractive index value at visible frequencies with a metamaterial fabricated using quartz as dielectric substrate and silver to form cut wire structure[26].

In 2011, M. Choi, et al., exquisitely combines both J. Shin[24] and H. Shi's[25] schemes to demonstrate high refractive index value at terahertz frequencies. The resonating subwavelength cut wire structure in H. Shi's scheme was replaced with an T-shaped metallic patch[27], an abridged 2D version of J. Shin's silted cube structure. The resultant metamaterial structure, as shown in Fig. 5(c) is capable of producing high effective permittivity due strong to the capacitive behavior while maintaining a low diamagnetic response due to the smaller current-loop area. M. Choi's metamaterial was fabricated on flexible, thin film polyimide substrate, the metallic structure was created using gold (on chromium), a periodic array of thus created unit cells is then layered on top of each other similar to H. Shi's scheme, to investigate the bulk properties of the metamaterial. The band structure and the dispersion relation of the fabricated metamaterial was investigated to understand the effect of number of layers on the electromagnetic characteristics of the bulk media. The maximized transmission of the slab is defined by the equation,

$$f_p = pc/2nNd, \quad (9)$$

where $f_p$ is the transmission peak frequency, $p$ is a non-negative integer, $N$ is the number of layers, $c$ is the speed of light and $d$ is the thickness of each layer[27]. M. Choi experimentally achieved a refractive index as high as 38.64 at 0.315 THz and claimed that higher values of refractive index can be achieved by reducing the distance between the individual T-shaped metallic patches or the spacing between each layer until the gap width reaches the Thomas–Fermi length scale[28] or the quantum tunneling scale of electrons.

The exact 3D structure of metallic cubes arranged in a cubic array fashion proposed by J. Shin, et al.[24], was experimentally demonstrated in microwave regime by T. Campbell, et al. in 2013[29]. T. Campbell fabricated two samples, one with array of solid copper cubes and another with an array of structured copper cubes Both the samples were characterized using strip line method for material characterization pioneered by W. Barry in 1986[30]. The comparison of the experimentally obtained material parameters of the solid cubes with that of the structured cubes revealed that the real part of permittivity of the structured cube sample was approximately half that of the solid cube sample but the refractive index was still nearly doubled since the structured cubes facilitated reduced diamagnetic response resulting in almost six-fold increase in the effective permeability value[29].

S. Tan, et al. furthered M Choi's T-shaped metallic patch structure[27] with a Z-shaped meta-atom in 2015[31], as shown in Fig. 5(d). The Z-shaped meta-atom was capable of attaining high refractive index at lower frequency by virtue of existence of a parallel, uni-directional surface current in the two arms of the structure which leads to larger effective induction as compared to that in I-structure thus lowers the resonance frequency without having to reduce the size the over-all structure. The Z-structure metamaterial was fabricated using conventional lithography using aluminum and thin film polyimide substrate. Time–domain terahertz spectroscopy characterization revealed 21.6% average shift in the resonance frequency[31]. This simple design modification offered great ease in fabrication of high refractive index metamaterials operating at lower frequencies. In 2016, T. Chang reported a broadband refractive index scheme based on mesoscopic space-filling curves. He claimed gigantic dielectric constant enhancement due displacement between the layers resulting in very high displacement enhancement factor[32].

Although these approaches were able to enhance refractive index to unnatural values, the highest values of refractive indices achievable with these schemes are demarcated mainly due to two reasons. First, the limit posed by the number of layer that can be productively used at a particular frequency, dictated by equation no. (8). Second, the practical limit to which the periodicity of the meta-atoms can be decreased, before losing the capacitance effect all together.

In 2017, we proposed a terahertz metamaterial design capable of achieving ever high values of refractive index owing to very high effective permittivity realized by employing deep subwavelength coupling between the arrays of identical structures situated on both sides of an ultra-thin dielectric film, as shown in Fig. 6[33].

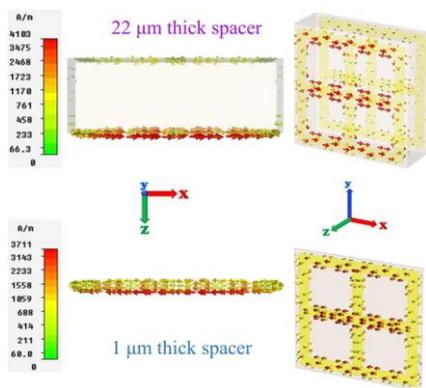

Fig. 6. Ultra-thin terahertz metamaterial with double sided metal structure. Depicting deep subwavelength coupling absence (above) and presence (below) between the metal structures situated on both sides of an ultra-thin dielectric substrate. *[Reprinted with permission from "L. Singh, R. Singh, and W. Zhang, Ultra-high terahertz index in deep subwavelength coupled bi-layer free-standing flexible metamaterials. J. Appl. Phys. 2017, 121(23): 233103", with the permission of AIP Publishing]*

Owing to the subwavelength thickness of the substrate film, the metal structures situated on both of its sides got coupled generating tremendous capacitive response in addition to the capacitance due to the nearest neighbor coupling between structures on the same side. Meanwhile the permeability of the metamaterial remained near unity due to small current loop area subtended by the checkboard metal structure. An array of isotropic checkboard structure was fabricated by thermal deposition of aluminum on both sides of polyimide substrate film of thickness 22 and 1 μm, respectively. Using terahertz-time domain spectroscopy, both the samples were characterized and the calculated effective refractive index for the 1 μm thick sample was found to be increased to 41.8 as compared to that of 7.78 refractive index for the 22 μm thick sample and can be further increased to 77.02 for a 100 nm thick sample. The enhancement of the refractive index can be further boosted by increasing deep subwavelength coupling between meta-atoms on the opposite side of the spacer layer either by decreasing the substrate layer thickness or selecting a substrate of higher refractive index or both. The effective refractive index of the metamaterial governed dominantly by the inter-layer deep subwavelength coupling, demonstrated a universal power law behavior (described by equation no. (10)) with respect to the substrate thickness, and a linear behavior (described by equation no. (11)) with respect to the substrate refractive index:

$$N_{effective} = \alpha t^\beta, \qquad (10)$$

$$N_{effective} = aN_{substrate} + b, \qquad (11)$$

where $\alpha, \beta, a$ and $b$ are the constants that depend on the metamaterial structure, $t$ is the substrate film thickness and $N_{substrate}$ is the substrate refractive index. Although it was predicted by many previous schemes[17,25, 27] that decreasing the thickness of individual designer layers of bulk metamaterial would enhance the refractive index values, it was the first time that reducing the substrate thickness to subwavelength levels was utilized to facilitate coupling effect between the resonating structures fabricated on both the sides of the substrate layer. This enabled to attain tremendous effective permittivity and thus ultra-high value of effective refractive index of the metamaterial. It is crucial to mention that although it is okay to achieve higher values of refractive index solely by enhancing the effective permittivity of the medium, the fact that permeability is also an integral part of the refractive index equation cannot be ignored. There have been a number of reports[34,35] on high refractive index metamaterial with miscalculated or unexplained permeability values above unity. The metamaterial structures discussed in these reports may be capable of refractive index enhancement as evident from the logical arguments presented in these reports but just not as high as reported, the miscalculated values are misleading and hence needs to be revisited for due corrections. Resonance based schemes[34,35] miscalculated the permeability to be higher than unity even in the absence of resonance. This is due to the fact that the measured transmission is not properly compensated to account for the experimental set up while plugging the measured values for calculating refractive index. Once calculations are set right, the resulting permeability values would be reduced, which would lead to a discrepancy in the reported refractive index values.

In conclusion, the science of designing high refractive index medium has come a long way since meddling with the intrinsic state of matter to the design of metamaterials with perpetually imaginative structures. It may not be even far-fetched to assume that the high refractive index media design with static values of refractive indices have quite saturated. In recent years the focus has shifted more towards devising media capable of dynamic enhancement of the refractive index[36,37] by means of electrical stimuli. The incredible scientific endeavors leading to realization of thin film, broadband, dynamic control, and ultra-high refractive index media are sure to play a crucial role in shaping the face of modern day devices.

The authors acknowledge R. Singh for contributions to this work.